\documentclass[epj,amsmath,amssymb,referee]{svjour} 
\usepackage{latexsym}
\usepackage{graphics}
\begin{document}
\title{Indications for the coexistence of closed orbit and quantum interferometer with the same cross section in the organic
metal
$\beta$''-(ET)$_4$(H$_3$O)[Fe(C$_2$O$_4$)$_3$]$\cdot$C$_6$H$_4$Cl$_2$:
Persistence of Shubnikov-de Haas oscillations above 30 K}
\subtitle{}
\author{David Vignolles\inst{1} \and Alain Audouard\inst{1}  \and  Vladimir N. Laukhin\inst{2,3}  \and Enric~Canadell\inst{3}
\and Tatyana G. Prokhorova\inst{4} \and Eduard B.
Yagubskii\inst{4}
}                     
\mail{audouard@lncmp.org}
%
\institute{Laboratoire National des Champs Magn\'{e}tiques
Intenses (UPR 3228 CNRS, INSA, UJF, UPS) 143 avenue de Rangueil,
F-31400 Toulouse, France. \and Instituci\'{o} Catalana de Recerca
i Estudis Avan\c{c}ats (ICREA), 08010 Barcelona, Spain. \and
Institut de Ci\`{e}ncia de Materials de Barcelona (ICMAB - CSIC),
Campus UAB, 08193 Bellaterra, Catalunya, Spain. \and Institute of
Problems of Chemical Physics, Russian Academy of Sciences, 142432
Chernogolovka, Moscow oblast, Russia.}
\date{Received: \today / Revised version: date}
%
\abstract{Shubnikov-de Haas (SdH) and de Haas-van Alphen (dHvA)
oscillations spectra of the quasi-two dimensional charge transfer
salt
$\beta$''-(ET)$_4$(H$_3$O)[Fe(C$_2$O$_4$)$_3$]$\cdot$C$_6$H$_4$Cl$_2$
have been investigated in pulsed magnetic fields up to 54 T. The
data reveal three basic frequencies F$_a$, F$_b$ and F$_{b - a}$,
which can be interpreted on the basis of three compensated closed
orbits at low temperature. However a very weak thermal damping of
the Fourier component F$_b$, with the highest amplitude, is
evidenced for SdH spectra above about 6 K. As a result,
magnetoresistance oscillations are observed at temperatures higher
than 30 K. This feature, which is not observed for dHvA
oscillations, is in line with quantum interference, pointing to a
Fermi surface reconstruction in this compound.
} 
\authorrunning{D. Vignolles et al.}
\titlerunning{Coexistence of closed orbit and quantum interferometer in $\beta$''-(ET)$_4$(H$_3$O)[Fe(C$_2$O$_4$)$_3$]$\cdot$C$_6$H$_4$Cl$_2$}
\maketitle
\section{Introduction}
\label{sec_intro}

The family of quasi-two-dimensional (q-2D) charge transfer salts
$\beta$''-(ET)$_4$(A)[M(C$_2$O$_4$)$_3$]$\cdot$Solv (where ET
stands for bis-ethylenedithio-tetrathiafulvalene, A is a
monovalent cation, M is a a trivalent cation and Solv is a
solvent) have raised great interest for many years \cite{Co04}.
Indeed, even though all the members of this family, denoted as (A,
M, Solv) hereafter, are isostructural, it is now established that
many different ground-states, including normal metal, charge
density wave, superconductivity, and temperature-dependent
behaviours can be observed \cite{Co04}.

\begin{figure}                                                      
\centering \resizebox{\columnwidth}{!}{\includegraphics*{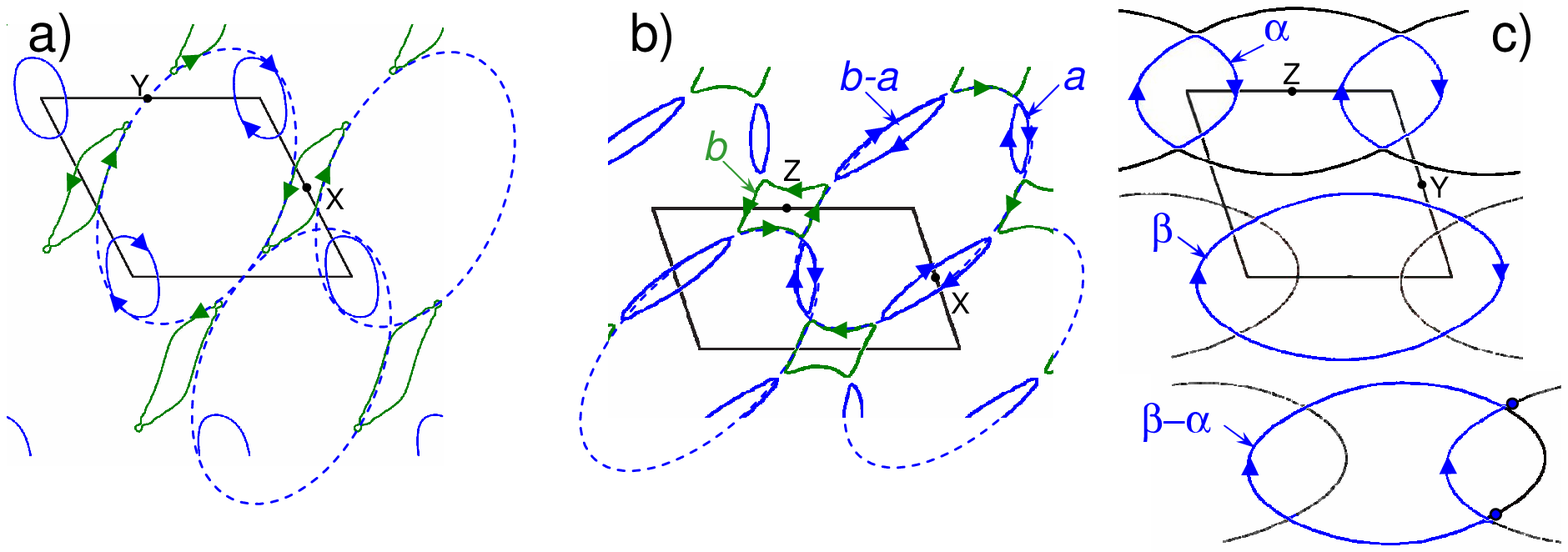}}
\caption{\label{fig_FS} (color on line) Fermi surface of (a)
$\beta$''-(ET)$_4$(NH$_4$)[Fe(C$_2$O$_4$)$_3$]$\cdot$C$_3$H$_7$NO
\cite{Pr03}, (b) (BEDO)$_4$Ni(CN)$_4\cdot$4CH$_3$CN \cite{Du05}
and (c) (BEDO)$_5$Ni(CN)$_4\cdot$3C$_2$H$_4$(OH)$_2$ \cite{Du05}.
Ellipses in dashed lines stand for intersecting hole orbits with
area equal to that of the First Brillouin zone ($\bigodot$
orbits). They lead to three compensated electron ($b$) and hole
($a$ and $b - a$) orbits in Fig. \ref{fig_FS}b. $\bigodot$ orbits
correspond to $\beta$ orbits of Fig. \ref{fig_FS}c. The arrows
indicate the quasiparticles direction (see text).}
\end{figure}

According to band structure calculations \cite{Pr03}, the Fermi
surface (FS) of these compounds originates from a hole orbit
(labelled $\bigodot$ in the following) with an area equal to that
of the first Brillouin zone (FBZ). Due to small gap opening,
compensated orbits with much smaller area (8.8 $\%$ of the FBZ,
according to \cite{Pr03}) are formed, as displayed in Fig.
\ref{fig_FS}a. Insofar as the Fermi level is very close to the
band extrema at Y \cite{Pr03,Ku95}, the actual FS topology can be
very sensitive to subtle structural details. Namely, in the case
where the $\bigodot$ orbits intersect along the $b^*$ direction,
an additional orbit is observed and the FS topology is similar to
that of Fig. \ref{fig_FS}b \cite{Du05}. Oppositely, if the gap is
larger than in the case of Fig. \ref{fig_FS}a, the electron-type
orbits transform into quasi-one dimensional sheets as displayed in
Fig. \ref{fig_FS}c \cite{Du05}. In addition, FS reconstruction due
to phase transition such as density wave condensation can further
modify the FS at low temperature. As a matter of fact, depending
on the studied compound, Shubnikov-de Haas (SdH) oscillations
spectra have revealed from one to six Fourier components in this
family \cite{Au04,Ba04,Col04,Ba05,Vi06}. Their frequency are in
the range 40 T to 350 T that corresponds to orbits with area
ranging from 2 to 18 $\%$ of the FBZ area. In most cases, these
orbits are compensated, in agreement with band structure
calculations. For example, SdH data of (NH$_4$, Fe, C$_3$H$_7$NO)
can be interpreted on the basis of three compensated orbits
\cite{Au04}, corresponding to the textbook case \cite{Ro04}
reported in Fig. \ref{fig_FS}b where $a$ and $b-a$ are hole orbits
while $b$ is an electron orbit. However, more complex SdH spectra,
that are strongly dependent on external parameters such as a
moderate applied pressure, can also be observed. In addition,
significant structural disorder, linked to the size of the solvent
molecule, has been reported for many members of this family
\cite{Tu99,Ra01,Ak02,Zo08}. Influence of the nature of the solvent
molecule on the physical properties, such as the behaviour of the
temperature dependence of the resistivity and the occurrence or
not of a superconducting ground state, has been considered. In
that respect, the temperature-dependent resistivity of (H$_3$O,
Fe, C$_6$H$_4$Cl$_2$) exhibits a metallic behaviour down to few K,
followed by a slight upturn. A magnetoresistance experiment
performed on this compound up to 17 T yielded a SdH spectrum
involving only two Fourier components \cite{Zo08}. The present
paper reports on both magnetoresistance and torque experiments
performed up to 54 T on this compound. A third Fourier component
is observed. Nevertheless, the main result is the persistence of
magnetoresistance oscillations at temperatures above 30 K. This
feature, which is not observed in de Haas-van Alphen (dHvA)
spectra is discussed on the basis of the presence of both a
quantum interferometer and a closed orbit with the same area. It
points to a FS reconstruction at low temperature.

\section{Experimental}

\label{sec_Exp} Magnetoresistance and magnetic torque were
measured in pulsed magnetic field up to 54 T with a pulse decay
duration of 0.32 s. A one-axis rotating sample holder allowed to
change the angle ($\theta$) between the direction of the magnetic
field and the \emph{c$^*$} crystal axis. For magnetoresistance
measurements, the studied crystal was an elongated hexagonal
platelet with approximate dimensions (0.4 $\times$ 0.2 $\times$
0.1)~mm$^3$, the largest faces being parallel to the conducting
\emph{ab}-plane. Electrical contacts were made using annealed
platinum wires of 20 $\mu$m in diameter glued with graphite paste.
Alternating current (10 $\mu$A, 77 Hz and 100 $\mu$A, 50 kHz for
zero-field and magnetoresistance measurements, respectively) was
injected parallel to the \emph{c$^*$} direction (interlayer
configuration). The explored temperature range was from 1.4 K to
32 K. Magnetic torque measurements were performed with a
commercial piezoresistive microcantilever \cite{Oh02} in the
temperature range from 1.9 K to 15 K. The crystal size was
approximately (0.3 $\times$ 0.1 $\times$ 0.07)~mm$^3$. Variations
of the cantilever piezoresistance was measured with a Wheatstone
bridge with an ac excitation at a frequency of 63 kHz \cite{Ja08}.
A lock-in amplifier with a time constant in the range 30 - 100
$\mu$s was used to detect the measured signal for
magnetoresistance and torque measurements. Discrete Fourier
analysis of oscillatory magnetoresistance and torque were
performed using Blackman-type window, which is known to avoid
secondary lobes.

\section{Results and discussion}
\label{sec_res}

\begin{figure}                                                      
\centering
\resizebox{\columnwidth}{!}{\includegraphics*{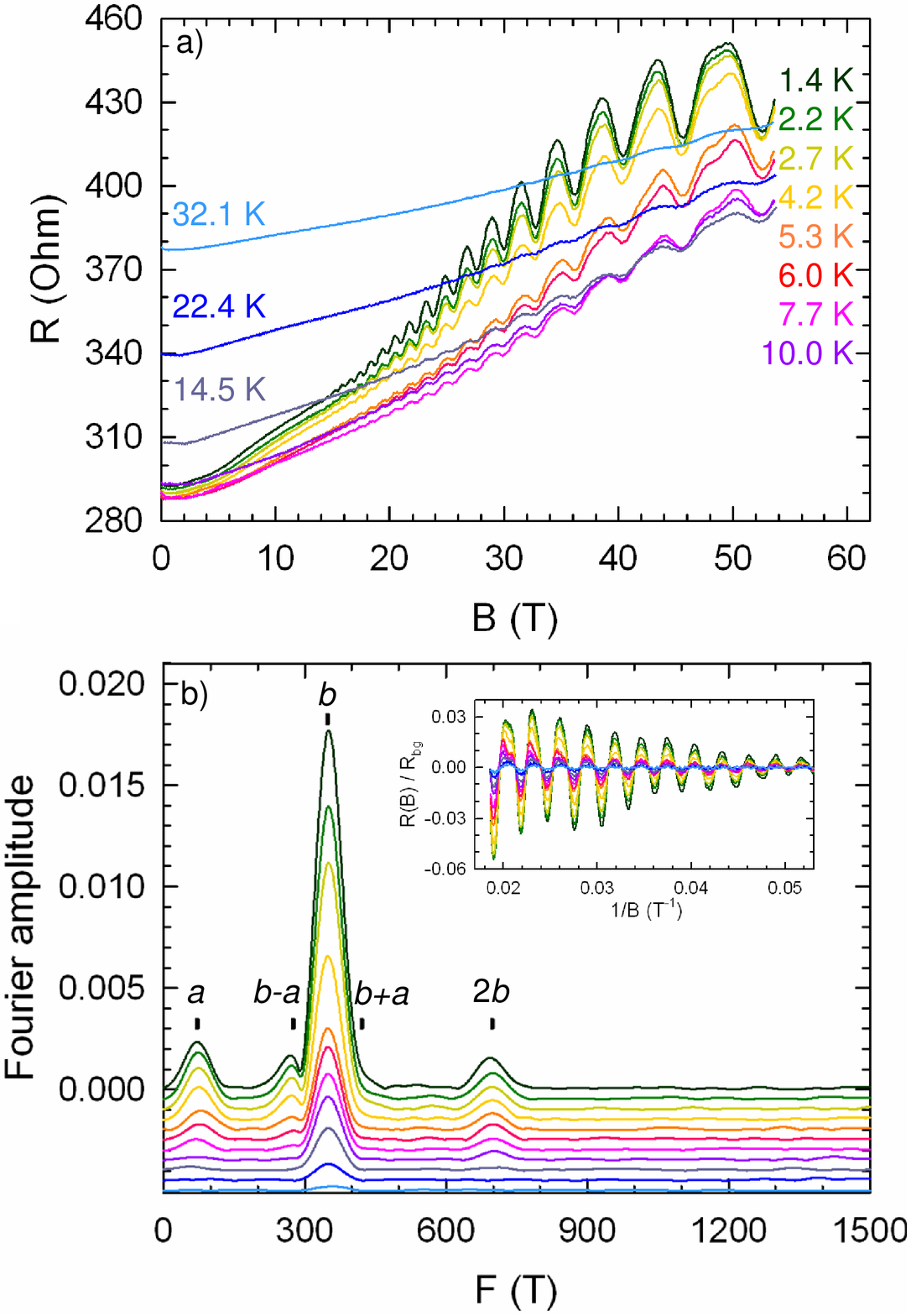}}
\caption{\label{fig_R(B)_0d} (color on line) (a) Field-dependent
interlayer resistance of
$\beta$''-(ET)$_4$(H$_3$O)[Fe(C$_2$O$_4$)$_3$]$\cdot$C$_6$H$_4$Cl$_2$
for $\theta$ = 0$^{\circ}$. (b) Fourier analyses deduced from the
oscillatory part of the magnetoresistance displayed in the inset.
The field range is 18-54 T. Marks are calculated with F$_a$ = 74 T
and F$_b$ = 348 T. }
\end{figure}

\begin{figure}                                                      
\centering
\resizebox{\columnwidth}{!}{\includegraphics*{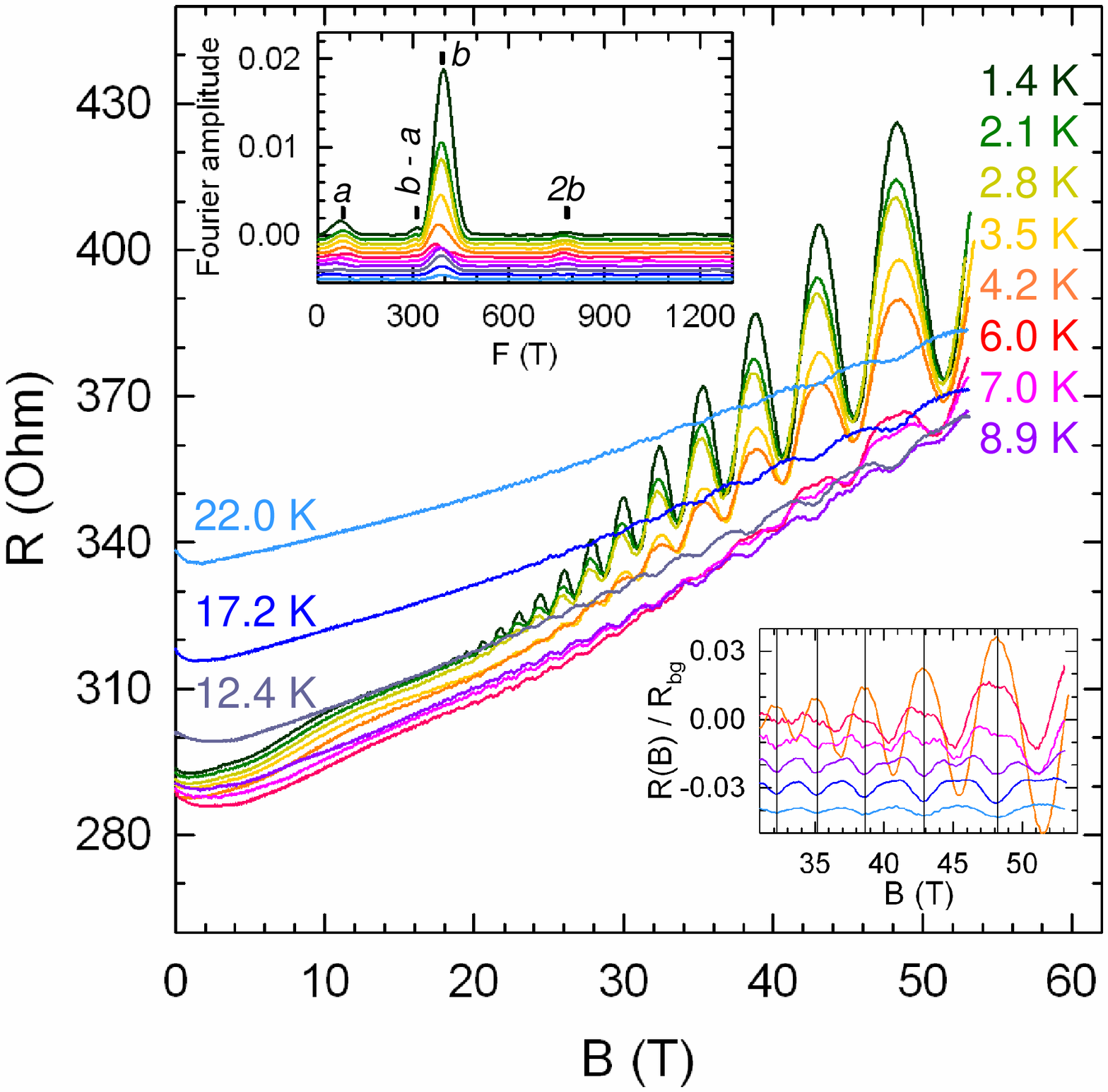}}
\caption{\label{fig_R(B)_27d} (color on line) Field-dependent
interlayer resistance for $\theta$ = 27$^{\circ}$. The upper inset
displays corresponding Fourier analyses. Oscillatory part of the
magnetoresistance data for temperatures above 4 K are displayed in
the lower inset to evidence the change of the Onsager phase factor
($\gamma_b$) discussed in the text. Vertical lines are marks
calculated with F$_b$ = 390 T and $\gamma$ = 0.1.}
\end{figure}


\begin{figure}                                                      
\centering
\resizebox{\columnwidth}{!}{\includegraphics*{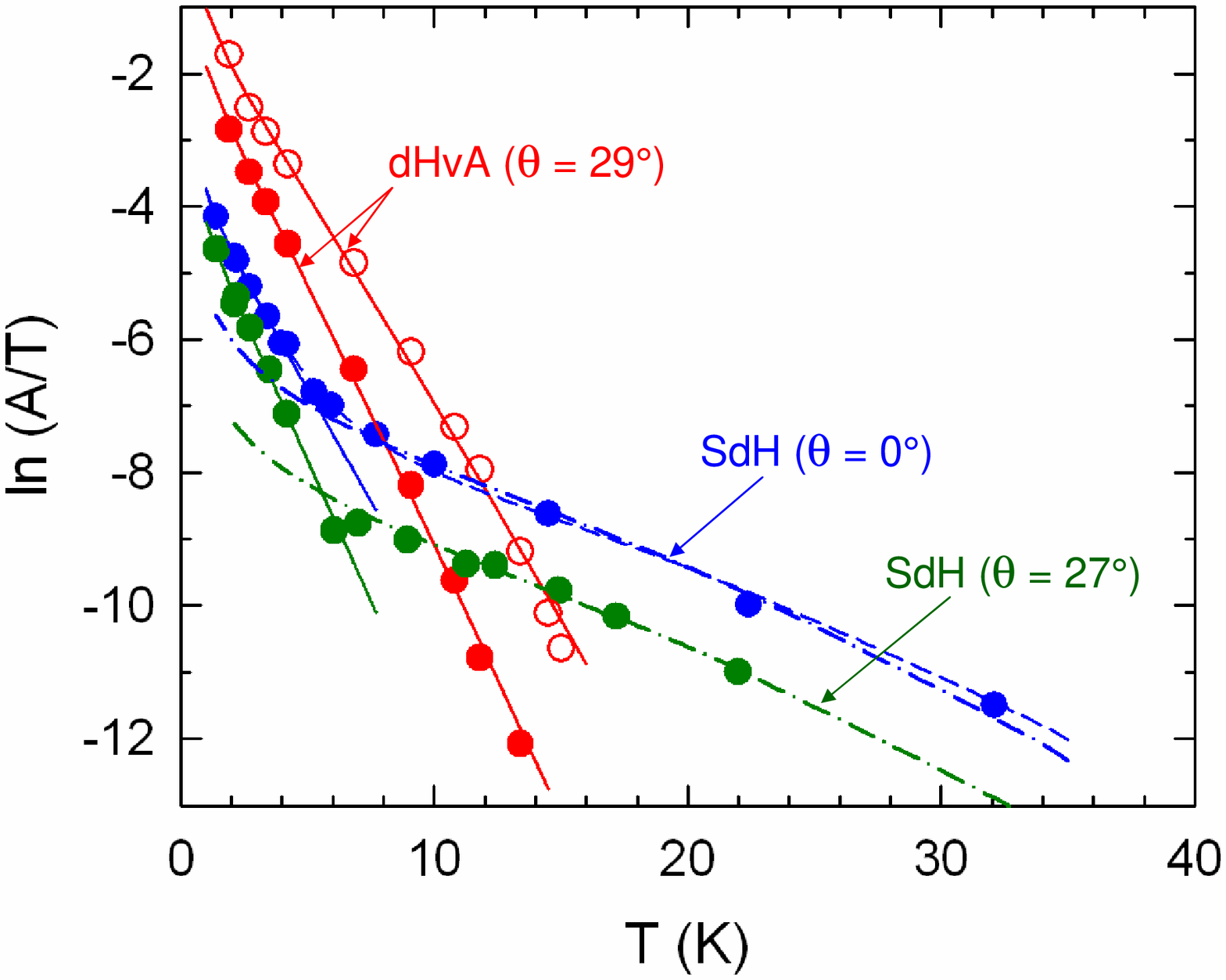}}
\caption{\label{fig_A(T)} (color on line) Temperature dependence
of the amplitude of the \emph{b} oscillations for dHvA and SdH
data. Empty and solid symbols correspond to a mean field value of
44.6 T and 30 T / cos($\theta$). SdH data for $\theta$ =
27$^{\circ}$ are shifted down for clarity. Solid and dash-dotted
lines are best fits of Eq. \ref{LK_SdH}, assuming the Dingle
damping factor is given by Eq. \ref{RD} and \ref{RD_hT},
respectively. In addition, a zero-effective mass is considered for
the SdH data in the high temperature range. Dashed line is a best
fit to the SdH data for $\theta$ = 0$^{\circ}$, assuming the
contributions of closed orbit and quantum interference coexists in
all the temperature range covered by the experiments (see text).}
\end{figure}

Fourier analysis of the oscillatory part of the magnetoresistance
data, obtained with magnetic field normal to the conducting plane
($\theta$ = 0$^{\circ}$), is displayed in Fig. \ref{fig_R(B)_0d}.
Four peaks can be identified with frequencies F$_a$ = 74  $\pm$ 5
T, F$_b$ = 348 $\pm$ 3 T, F$_{b-a}$ = 272 $\pm$ 5 T and F$_{2b}$ =
699 $\pm$ 5 T that accounts for the relationship F$_a$ + F$_{b-a}$
= F$_b$. The reported uncertainties are deduced from data below
2.2K for various field directions which allow checking that these
frequencies follow the cos($\theta$) dependence expected for a 2D
FS (up to $\theta$ = 62$^{\circ}$ for the $b$ component). F$_a$
and F$_b$ could correspond to the frequencies F$_2$ $\approx$ 80 T
and F$_1$ = 376 T, respectively, reported in \cite{Zo08}.
Oscillatory spectra including frequencies F$_a$, F$_b$ and their
combinations have already been observed in few other compounds of
the considered family \cite{Au04,Vi06,Au06} and interpreted on the
basis of a FS similar to that displayed in Fig. \ref{fig_FS}b
\cite{Du05}, taking into account that the respective location of
the hole orbits $a$ and $b-a$ can be exchanged. It is well known
that linear relationship between frequencies can also be observed
for FS topology analogous to Fig. \ref{fig_FS}c, namely
F$_{\alpha}$ + F$_{\beta - \alpha}$ = F$_{\beta}$, where $\beta$ -
$\alpha$ is a QI path. However, the area of the MB-induced $\beta$
orbit is equal to that of the FBZ. In the case of the compound of
Fig. \ref{fig_FS}c, F$_{\beta}$ = 3837 T \cite{Vi07}. Such a
frequency value would be close to that expected for the $\bigodot$
orbit (F$_{\bigodot}$ = 3975 T) which is an order of magnitude
larger than F$_b$.

In the framework of the Lifshits-Kosevich (LK) formula, and
assuming that the amplitude of the oscillations is small compared
to the background non-oscillating part of the resistance
(R$_{bg}$), the oscillatory magnetoresistance (R(B)/R$_{bg}$) can
be accounted for by:

\begin{eqnarray}
\label{LK_SdH} \frac{R(B)}{R_{bg}} = 1 + \sum_{j}A_j
cos[2{\pi}(\frac{F_j}{B}-\gamma_j)]
\end{eqnarray}

where F$_j$ and $\gamma_j$ are the frequency and the phase factor,
respectively, of the Fourier component linked to the orbit $j$.
The Fourier amplitude is given by A$_j \propto$
R$_{Tj}$R$_{Dj}$R$_{MBj}$R$_{Sj}$. The thermal (for a 2D FS),
Dingle, magnetic breakdown (MB) and spin damping factors are
respectively given by \cite{Sh84}:

\begin{eqnarray}
\label{RT} R_{Tj} =
\frac{{\alpha}Tm{_j^*}}{Bsinh[{\alpha}Tm{_j^*}/B ]}
\\\label{RD}R_{Dj} =
exp[-{\alpha}T_{D}m{_j^*}/B]\\
\label{RMB}R_{MBj} =
exp(-\frac{t_jB_{MB}}{2B})[1-exp(-\frac{B_{MB}}{B})]^{b_j/2}\\\label{RS}R_{Sj}
=cos(\pi\mu_j/ cos \theta)
\end{eqnarray}

where  m$_j^*$ is the effective mass normalized to the free
electron mass m$_e$, T$_{D}$ = $\hbar\tau^{-1}$/2$\pi$k$_B$ is the
Dingle temperature, $\tau^{-1}$ is the scattering rate, $\mu_j =
g^*m_j^*(\theta = 0)$/2, $g^*$ is the effective Land\'{e} factor
and B$_{MB}$ is the MB field. Integers $t_j$ and $b_j$ are
respectively the number of tunnelling and Bragg reflections that
the quasiparticles come up against along their path.

In addition to Fourier analysis, information regarding the
oscillatory spectra can also be derived by direct fitting of the
LK formula to the field-dependent magnetoresistance. Indeed,
direct fitting is useful in order to discern eventual Fourier
components with close frequencies and allow for a reliable
determination of $\gamma_j$. In the high T/B range, and assuming
t$_j$ = 0 (see Eq. \ref{RMB}), Eq. \ref{LK_SdH} can be
approximated to:

\begin{equation}
\label{LK_approx} \frac{R(B)}{R_{bg}} = 1 + \sum_{j}
\frac{a_j}{B}exp(-\frac{B_j}{B})
cos[2{\pi}(\frac{F_j}{B}-\gamma_j)]
\end{equation}

where a$_j$ is a prefactor.  Eq. \ref{LK_approx} involves a
reduced number of free parameters compared to Eq. \ref{LK_SdH}
since B$_j$ $\simeq$ $\alpha$(T + T$_{Dj}$)m$_i^*$. Its drawback
is that, strictly speaking, it is only valid at high temperature,
high T$_D$ and (or) low field. Nevertheless, it is only intended
to identify the various Fourier components entering the data and
to derive the parameters F$_j$ and $\gamma_j$. In that respect,
Eq. \ref{LK_approx} has been successfully used at low temperature
and high field for SdH oscillations of (NH$_4$, Cr, DMF) under
pressure \cite{Vi06} and dHvA oscillations of high-T$_c$
superconductors \cite{Ja08}. In the present case, a very good
agreement between frequencies deduced from direct fittings and
Fourier analysis is obtained.

Turn on now to the most striking result evidenced in Figs.
\ref{fig_R(B)_0d} and \ref{fig_R(B)_27d}, namely the persistence
of quantum oscillations at temperatures higher than 30 K (up to 22
K at $\theta$ = 27$^{\circ}$). More insight on this feature can be
derived from the field and temperature dependence of the Fourier
components observed in the oscillatory part of the
magnetoresistance.

\begin{figure}                                                      
\centering \resizebox{\columnwidth}{!}{\includegraphics*{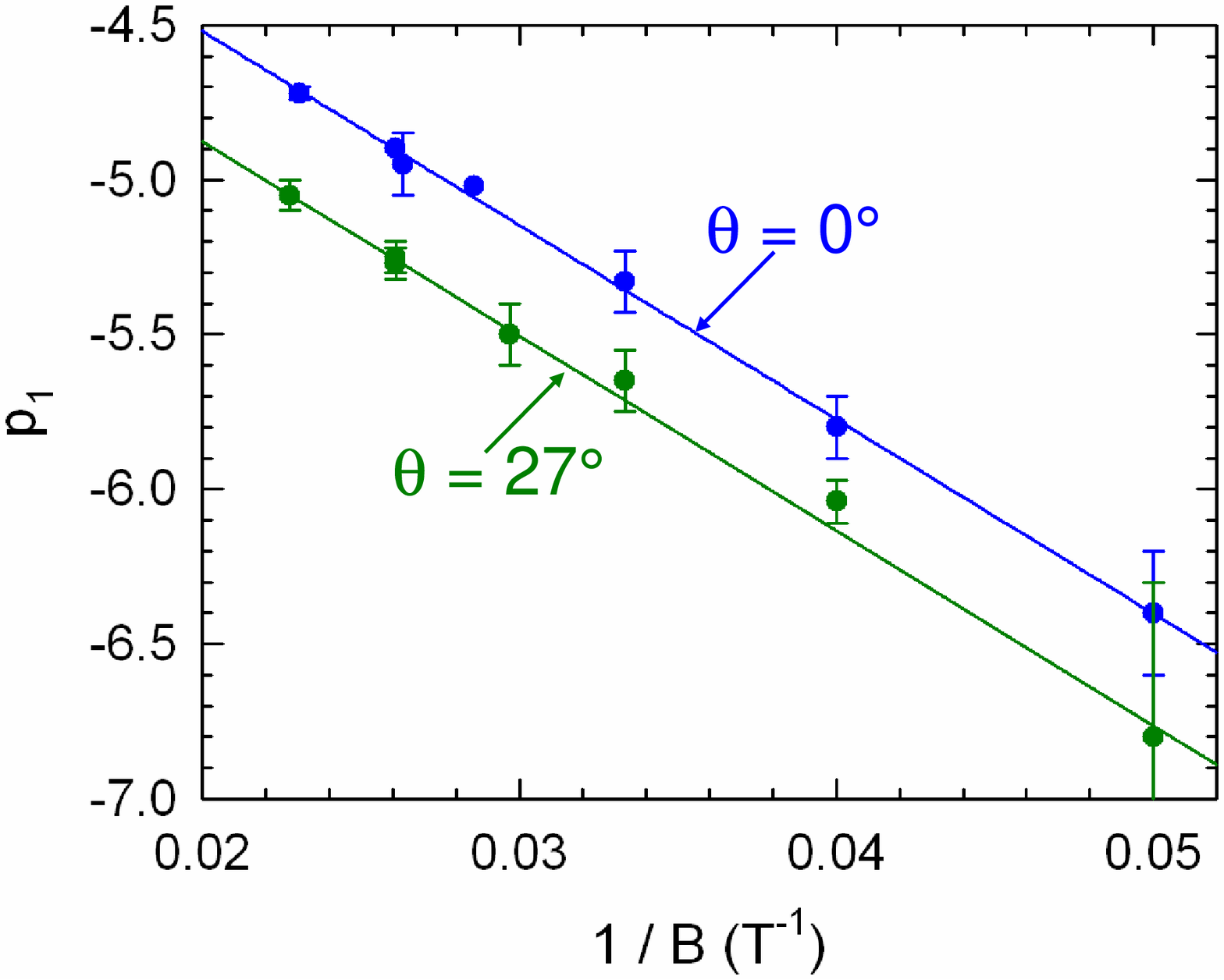}}
\caption{\label{fig_p3} (color on line) Field dependence of the
parameter p$_1$ = a$_0$ - $\alpha$m'T$_{D0}$/B, deduced from the
temperature dependence of the \emph{b} oscillations amplitude in
the high temperature range for $\theta$ = 0$^\circ$ and
27$^\circ$. Solid straight lines are best fits to the data (see
text).}
\end{figure}

As usual, effective masses can be derived from the temperature
dependence of the oscillations amplitude through Eq. \ref{RT}.
Data at $\theta$ = 0$^{\circ}$ yield  m$^*_a$ = 0.85 $\pm$ 0.10
and m$^*_{b-a}$ = 1.0 $\pm$ 0.1 for $a$ and $b - a$ Fourier
components, respectively. Data for the $b$ oscillations are
displayed in Fig. \ref{fig_A(T)}. At low temperature, they yield
m$^*_b$ = 1.4 $\pm$ 0.2. The three effective mass values given
above lie in the same range as for other compounds of this family.
However, a clear kink can be observed in Fig. \ref{fig_A(T)} as
the temperature increases above $\sim$ 6 K. The temperature
dependence of the amplitude is much less steep in the high
temperature range which accounts for the observation of
magnetoresistance oscillations at temperatures as high as 32 K in
Fig. \ref{fig_R(B)_0d}. The slope of the mass plot would yield m*
$\simeq$ 0.35 at high temperature. However, inelastic collisions
cannot be neglected in this range. Indeed, the zero-field
interlayer resistance starts to significantly increase above a few
kelvins \cite{Zo08}. As a result, inelastic collisions should
account, at least for a part, for the damping of the oscillation
amplitude at high temperature. Therefore, the effective mass
should be even smaller than the above value. Actually, the data
are compatible with a zero effective mass, such as that coming
from a symmetric quantum interferometer \cite{St71,St74,Vi03}.
Indeed, assuming the decrease of the \emph{b} oscillation
amplitude is entirely due to the increase of the inelastic
scattering rate $\tau_i^{-1}$ as the temperature increases and
assuming further $\tau^{-1}$ = $\tau_0^{-1}$ + $\tau_i^{-1}$ and
$\tau_i^{-1} \propto$ T$^2$, allows to rewrite the Dingle damping
factor:

\begin{equation}
\label{RD_hT}                                                        
R_{D} = exp[-\frac{{\alpha} m'}{B}(T_{D0} + \beta T^2)]
\end{equation}

where m' is the sum of the effective masses linked to each arms of
the interferometer \cite{St71,St74,Vi03} and $\beta$ is a
prefactor. In this case, still assuming a quantum interferometer
with a zero effective mass and neglecting MB, reduces the
contribution to the \emph{b} oscillations amplitude (A$_b$) in the
high temperature range to the temperature-dependent R$_D$ given by
Eq. \ref{RD_hT}. This leads to A$_b$ = exp(p$_1$ + p$_2$T$^2$)
where p$_1$ = a$_0$- $\alpha$m'T$_{D0}$/B (a$_0$ is a constant).
Dash-dotted lines in Fig. \ref{fig_A(T)} are best fits to the data
in which the product $\alpha$m'$\beta$ has been fixed to 0.085
TK$^{-2}$. In agreement with Eq. \ref{RD_hT} which assumes a
zero-effective mass, p$_1$ decreases linearly with 1/B (see Fig.
\ref{fig_p3}), even though a zero-effective mass must be regarded
as the lowest limit. The value of the slope, which is negligibly
dependent on the value of assumed for $\alpha$m'$\beta$
\footnote{A change of the $\alpha$m'$\beta$ value, i.e. of p$_2$,
produces a field-independent shift of the deduced p$_1$ value. In
addition, since a T$^2$ behaviour of the interlayer resistance is
only observed in the range 10-20K, we have checked that assuming
$\tau_i \propto$ T in Eq. \ref{RD_hT} still satisfactorily
accounts for the data.}, yields m'T$_{D0}$ = 4.3 $\pm$ 0.2 K.

\begin{figure}                                                      
\centering
\resizebox{\columnwidth}{!}{\includegraphics*{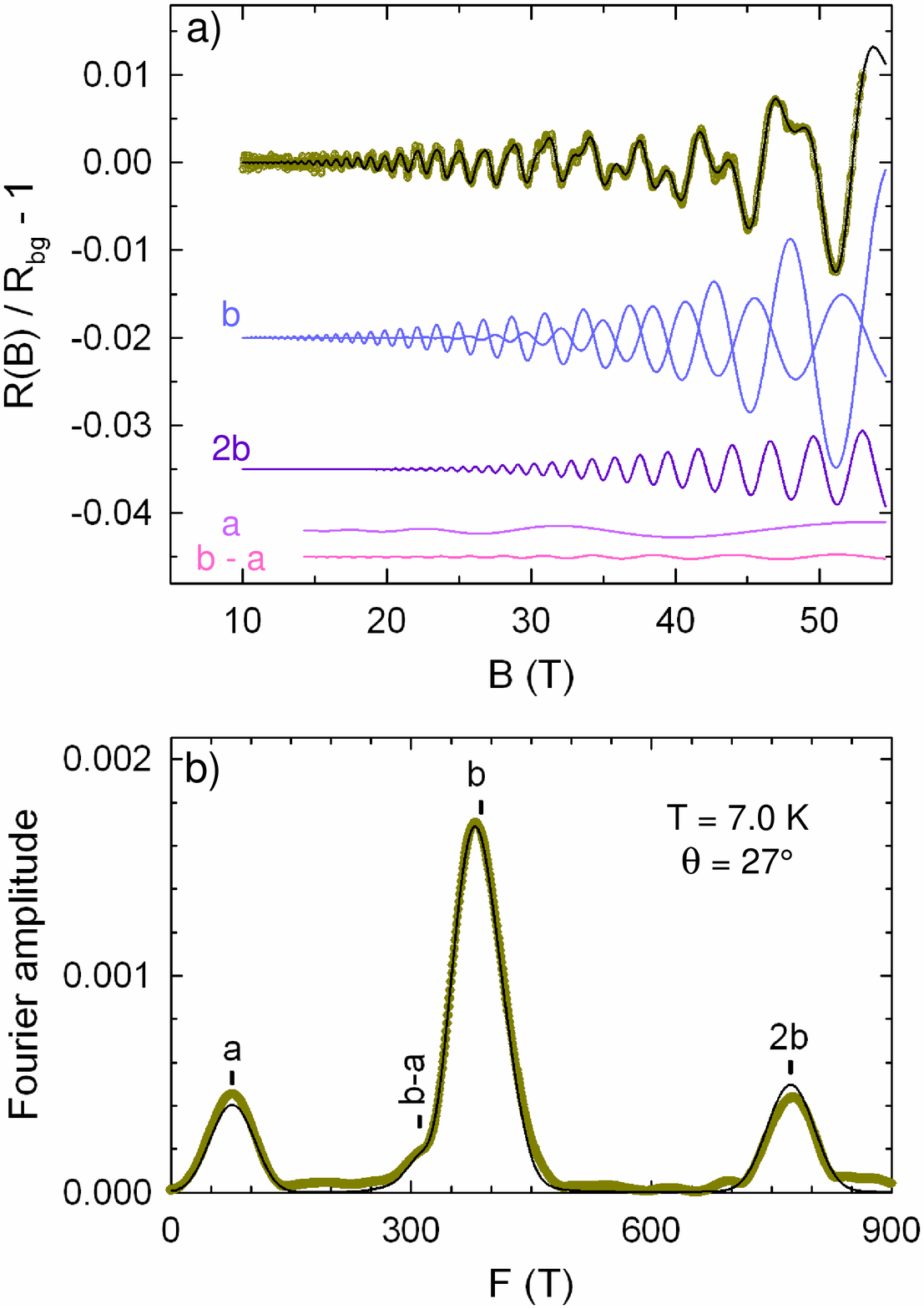}}
\caption{\label{fig_fit_m_TF_27d} (color on line) (a) Best fit of
Eq. \ref{LK_approx} (black solid line) to the oscillatory
magnetoresistance (yellow solid symbols) at 7.0 K and $\theta$ =
27 $^{\circ}$. The various components entering the fittings are
shifted from each other by an arbitrary amount. (b) Fourier
analysis of the experimental data (yellow solid symbols) and of
the best fit (black solid line). The field range is 18 - 53.5 T.
Marks correspond to the frequency of the various components
entering the fittings. They are obtained with F$_a$ = 76.6 T and
F$_b$ = 386.9 T.}
\end{figure}

\begin{figure}                                                      
\centering
\resizebox{\columnwidth}{!}{\includegraphics*{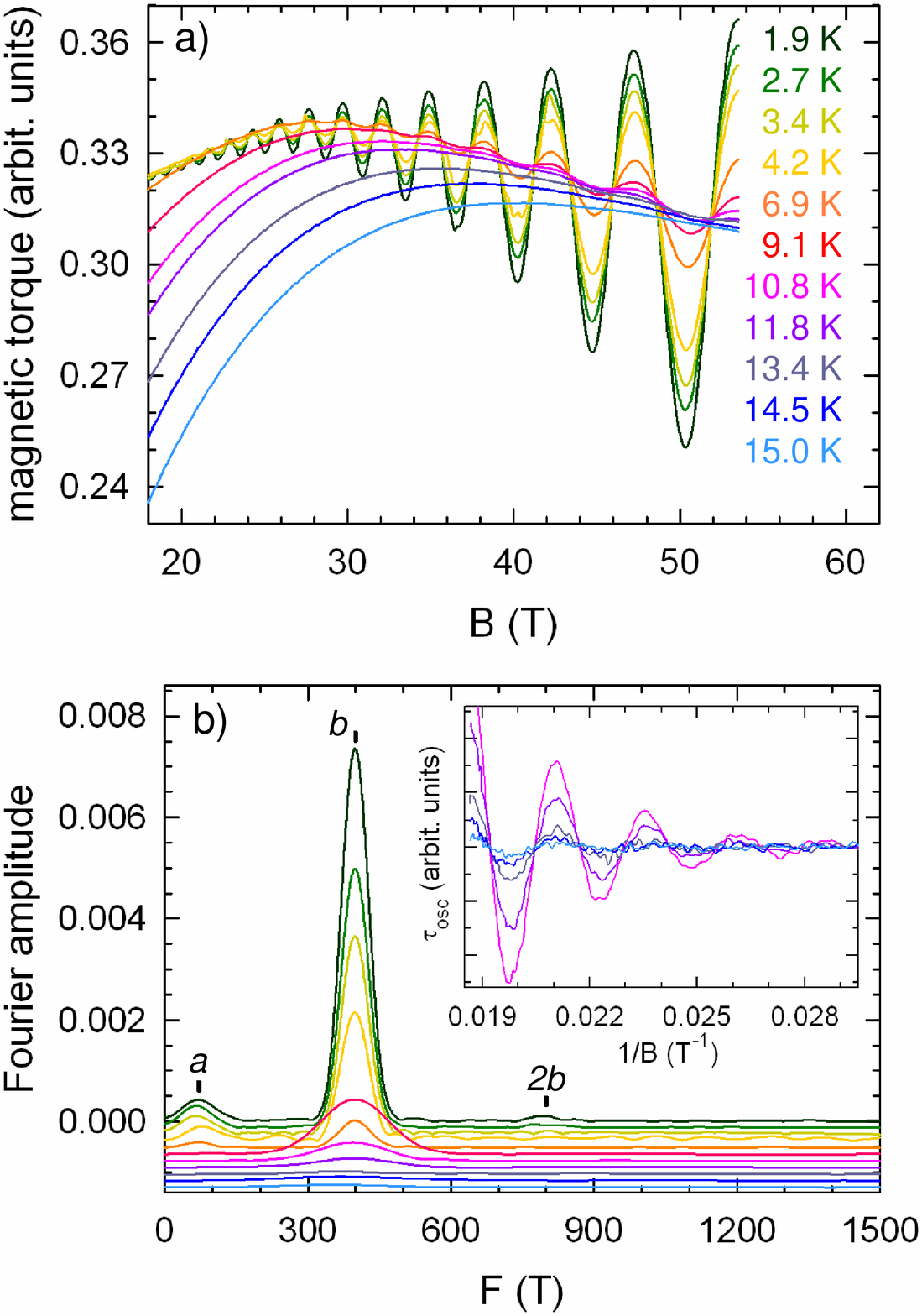}}
\caption{\label{fig_torque} (color on line) (a) Magnetic torque of
$\beta$''-(ET)$_4$(H$_3$O)[Fe(C$_2$O$_4$)$_3$].DCB at $\theta$ =
29$^{\circ}$. (b) Fourier analyses deduced from data in (a). The
field range is 30-53.5 T and 38-53.5 T below and above 9 K,
respectively. The inset displays the oscillatory part of the data
in (a) for temperatures above 10 K.}
\end{figure}

In the above discussion, it is implicitly assumed that each of the
two phenomena (either closed orbit or quantum interference)
involved in the $b$ Fourier component occurs in a separate
temperature range or, at the very least, is strongly dominant in
the considered temperature range. Oppositely, coexistence of the
two phenomena can be directly observed in the temperature range
$\sim$ 6 - 9 K for $\theta$ = 27$^{\circ}$. Indeed, in this range,
two out of phase oscillations series with the frequency F$_b$ can
be distinguished in the oscillatory magnetoresistance (see the
lower inset of Fig. \ref{fig_R(B)_27d}). More insight on this
behaviour can be obtained thanks to direct fittings of Eq.
\ref{LK_approx} that reveal the two $b$ components with phase
factors shifted by about 1/2, as can be observed in Fig.
\ref{fig_fit_m_TF_27d}. More precisely, direct fittings in the
considered temperature range yield $\gamma_b$ = 0.1 $\pm$ 0.1 and
0.55 $\pm$ 0.10, respectively. In contrast, only one component,
with $\gamma_b$ = 0.15 $\pm$ 0.10, can be observed in all the
temperature range studied for $\theta$ = 0. This behaviour is in
line with the angle dependence of the spin zero damping factor
(see Eq. \ref{RS}) for the $b$ component at low temperature.
Indeed, a good agreement between the angle dependence of the $b$
component amplitude at 2 K and the LK formalism is observed. It
yields $\mu$ = 1.6 $\pm$ 0.1, in rough agreement with the m$_b$
value deduced from the low temperature part of the data, which
suggests that g$^*$ is close to 2. Therefore, not any zero can be
observed below $\theta$ = 50$^{\circ}$. The dephasing by about
$\pi$ in the high temperature range for $\theta$ = 27$^{\circ}$
could then be ascribed to a peculiar angle dependence of the QI
phenomenon. Going further, it can be considered that the quantum
interferometer responsible for the $b$ component above $\sim$ 6 K
is already present at the lowest temperature explored. In such a
case, the Fourier amplitude should be the sum of the amplitudes
linked to the SdH (A$_{SdH}$) and QI (A$_{QI}$) phenomena in all
the temperature range studied, provided the phase factors are the
same\footnote{Assuming the contribution of the $b$ component to
the oscillatory magnetoresistance (Y$_b$) is given by Y$_b$ =
A$_{SdH}$cos[2$\pi$(F$_b$/B+$\gamma_{SdH}$)] +
A$_{QI}$cos[2$\pi$(F$_b$/B+$\gamma_{QI}$)] yields Y$_b$ =
A$_b$cos[2$\pi$(F$_b$/B+$\gamma_b$)] with A$_b^2$ = A$_{SdH}^2 +
$A$_{QI}^2$ +
2A$_{SdH}$A$_{QI}$cos[2$\pi$($\gamma_{SdH}$-$\gamma_{QI}$)] and
tan$\gamma_b$ = [A$_{SdH}$sin(2$\pi$$\gamma_{SdH}$) +
A$_{QI}$sin(2$\pi$$\gamma_{QI}$)]/[A$_{SdH}$cos(2$\pi$$\gamma_{SdH}$)
+ A$_{QI}$cos(2$\pi$$\gamma_{QI}$)]. } which is the case for
$\theta$ = 0$^{\circ}$, as discussed above. The dashed line in
Fig. \ref{fig_A(T)} is the corresponding best fit of Eq.
\ref{LK_SdH} to the data: a very good agreement is observed which
indicates that the above hypothesis can be considered. The deduced
effective mass is m$_b^*$ = 1.6 $\pm$ 0.1 which is in excellent
agreement with the above reported value of $\mu$.

The above data analysis suggests that a QI phenomenon accounts for
the temperature dependence of the $b$ component amplitude. In the
case where this hypothesis is valid, magnetization which as a
thermodynamic parameter is only sensitive to the density of states
should not reveal any contribution of QI. Torque data obtained for
a field direction $\theta$ = 29$^{\circ}$ are displayed in Fig.
\ref{fig_torque}(a). It can be remarked that F$_{b-a}$ is not
detected in dHvA spectra which could suggests that it corresponds
to a QI path, as well, even though its effective mass is rather
large (m$^*$ = 1.0 $\pm$ 0.1, see above). However, owing to the
$\mu_i$ value deduced from the angle dependence of the $b$ SdH
oscillations, this field direction yields the largest amplitude
for this component. Oppositely, the $b-a$ oscillations amplitude
is strongly reduced, as evidenced in Fig. \ref{fig_R(B)_27d} for
$\theta$ = 27$^{\circ}$. Namely, the ratio A(b-a)/A(b) is
decreased by a factor of 5 as $\theta$ goes from 0 to
29$^{\circ}$. In addition, dHvA oscillations are detected in a
narrower field range than for SdH oscillations which broadens the
Fourier peaks. Actually, only the $a$, $b$, and 2$b$, components
can be detected in the Fourier analysis of the oscillatory part of
the data (see Fig. \ref{fig_torque}(b)). In the framework of the
LK theory, the oscillatory part of the magnetic torque
$\tau_{osc}$ can be expressed as \cite{Sh84,Wo96}:

\begin{eqnarray}
\label{LK_torque}                                                  
\tau_{osc} = B tan \theta
\sum_{j}T_jsin[2{\pi}(\frac{F_j}{B}-\gamma_j)],
\end{eqnarray}

where T$_j \propto$ R$_{Tj}$R$_{Dj}$R$_{MBj}$R$_{Sj}$. Examples of
temperature dependence of the $b$ oscillations amplitude are
reported in Fig. \ref{fig_A(T)}. As displayed in Fig.
\ref{fig_torque}(b), oscillations are detectable up to 15 K, i. e.
at temperatures much higher than 6 K, at which QI oscillations
start to mostly contribute to the magnetoresistance oscillations
amplitude. In contrast to SdH data, no kink is observed for dHvA
oscillations. This behaviour corroborates the assumption that QI
is responsible for the persistence of magnetoresistance
oscillations at high temperature. The effective mass deduced from
dHvA data in Fig. \ref{fig_A(T)} corresponds to m$^*_b$($\theta$ =
0$^{\circ}$) = 1.5 $\pm$ 0.1 which is in good agreement with SdH
data at low temperature. The question remaining to solve deals
with the actual FS topology allowing both closed orbit and QI with
the same frequency. In that respect, no decisive clue is observed
since the main feature that could eventually account for FS
reconstruction is a slight upturn of the zero-field resistance at
low temperature \cite{Zo08}.

The Dingle temperature deduced from dHvA data, T$_D$ = 3.3 $\pm$
0.5 K, is roughly a factor of two lower than for (NH$_4$, Fe,
C$_3$H$_7$NO) and (NH$_4$, Cr, C$_3$H$_7$NO) \cite{Au04,Vi06}.
Even though T$_D$ can be significantly crystal dependent, it can
be remarked that this value is in the same range as for (H$_3$O,
M, C$_5$H$_5$N), where M = Ga, Cr, Fe \cite{Col04}. These rather
high values are in line with the significant structural disorder
observed in most of these compounds \cite{Tu99,Ra01,Ak02}.
According to Ref. \cite{Tu99} the disorder level is related to the
ability of the solvent molecule to fill the cavities of the anion
layer and is therefore linked to the solvent molecule size. More
precisely, it has been inferred that the length of the solvent
molecule along the $b$ crystallographic axis is the pertinent
parameter for the reduction of both the disorder level and the
increase of the superconducting transition temperature
\cite{Zo08}. However, owing to the rather poor correlation between
the reported T$_D$ values and the solvent molecule size (either
volume or length), it can be inferred that this latter parameter
cannot alone account for the scattering rate deduced from quantum
oscillations.

\section{Summary and conclusion}
SdH and dHvA oscillations spectra of the q-2D charge transfer salt
$\beta$''-(ET)$_4$(H$_3$O)[Fe(C$_2$O$_4$)$_3$]$\cdot$C$_6$H$_4$Cl$_2$
reveal three main frequencies F$_a$ = 74  $\pm$ 5 T, F$_{b-a}$ =
272 $\pm$ 5 T and F$_b$ = 348 $\pm$ 3 T. These frequencies are
linked through the linear combination F$_a$ + F$_{b-a}$ = F$_b$
that is consistent with a FS composed of three compensated closed
orbits, as already reported for few other members of this family
\cite{Au04,Vi06,Au06}. Effective masses, m$_a^*$ = 0.85 $\pm$
0.10, m$_{b-a}^*$ = 1.0 $\pm$ 0.1 and m$_b^*$ = 1.5 $\pm$ 0.1, are
in the same range as for the other salts of the family
\cite{Au04,Ba04,Col04,Ba05,Vi06,Au06}. However, as for SdH
spectra, the temperature dependence of the $b$ component amplitude
exhibits a kink at about 6 K, followed by a very weak variation.
As a result, oscillations can be observed up to 32 K. This feature
is not observed for dHvA oscillations (up to 15 K) which is
consistent with a QI phenomenon arising from a quantum
interferometer with a zero effective mass, having the same cross
section as the closed orbit responsible for the $b$ oscillations
in the low temperature range. Both of them coexist, at least in
the range 6 - 9 K for which two Fourier components with the
frequency F$_b$, roughly in phase opposition, can be observed for
$\theta$ = 27$^{\circ}$, likely due to a different spin-dependent
behaviour of the two contributions. In fact, coexistence of these
two phenomena in all the temperature range studied can be
considered. This result would be in line with the hypothesis of a
quantum interferometer built on the $b$ closed orbit thanks to FS
reconstruction at low temperature. Unfortunately, in the absence
of a reliable knowledge of the FS topology at low temperature,
this conclusion must be taken with caution.

Finally, the deduced Dingle temperature lies within the range of
the values reported so far in the literature for the compounds of
this family. No clear dependence of the scattering rate on the
solvent molecule size can be inferred from the available data.

\begin{acknowledgement}
 This work has been supported by  FP7 I3 EuroMagNET II and by the French-Spanish exchange programm
between CNRS and CSIC (number 18858).
 \end{acknowledgement}




\end{document}